\documentclass[draft,jgrga]{agu2001}
\usepackage{lineno}
\linenumbers*[1]
\usepackage{graphicx}
\authorrunninghead{YERMOLAEV}

\titlerunninghead{COMMENT}

\authoraddr{Yu. I. Yermolaev
Space Plasma Physics Department, Space Research Institute, 
Russian Academy of Sciences, Profsoyuznaya 84/32, Moscow 117997, Russia. 
(yermol@iki.rssi.ru)} 

\begin{document}

%
%
%
%
%

%
%

\title{Comment on "Geoeffectiveness of halo coronal mass ejections" by Gopalswamy, 
       N., S. Yashiro, and S. Akiyama"  
}

%
%



\author{Yu. I. Yermolaev} 
\affil{Space Plasma Physics Department, Space Research Institute, 
Russian Academy of Sciences, Profsoyuznaya 84/32, Moscow 117997, Russia}

\begin{abstract}
(No abstract for comment)
\end{abstract}

%
%

%

\begin{article}

%
%


\citet{Gopalswamy2007}
studied the geoeffectiveness of halo coronal mass ejections (CMEs) 
on the basis of solar observations during 1996-2005 and found that 
the geoeffectiveness of 229 frontside halo CMEs was 71\%. Recently 
for observations of 305 frontside halo CMEs during 1997-2003 the 
geoeffectiveness was found to be 40\% 
\citep{Kimetal2005}. 
Complex analysis of both solar and interplanetary measurements showed 
that the geoeffectiveness of frontside halo CMEs is likely to be about 50\% 
\citep{Yermolaevetal2005,Yermolaev2006}.
\citet{Gopalswamy2007}
did not discuss possible causes of this difference and were limited only 
to the general words: "The reason for the conflicting results 
(geoeffectiveness of CMEs ranging from 35\% to more than 80\%) may be 
attributed to the different definition of halo CMEs and geoeffectiveness". 
So, here we shall present our point of view on high geoeffectivenees 
of CME obtained in paper by 
\citet{Gopalswamy2007}. 

Different statistics of frontside halo CMEs (305 events during 1997-2003 
\citep{Kimetal2005} 
and 229 events during 1996-2005 
\citep{Gopalswamy2007}) indicates that 
\citet{Gopalswamy2007} 
used harder criteria of event selection. They wrote: "The solar source of 
a halo CME is usually given as the heliographic coordinates of any 
associated eruption region obtained in one or more of the following ways: 
(1) using H-alpha flare location if available from the Solar Geophysical 
Data, (2) running EIT movies with superposed LASCO images to identify any 
associated disk activity such as EUV dimming, and (3) identifying the 
centroid of the post eruption arcades in X-ray and EUV images when available" 
and then "For backside halos we do not see any disk activity". 
On the other hand, attempts of association of interplanetary CMEs (ICMEs) 
with coronal CMEs showed that this approach is incorrect. 
\citet{Zhangetal2007}
wrote in Introduction: "a number of ICMEs, including those causing major 
geomagnetic storms, were found not to be associated with any identifiable 
frontside halo CMEs 
\citep{Zhangetal2003,Schwennetal2005}". 
They studied sources of 88 large magnetic storms (Dst $<$ -100 nT) during 
1996-2005 and found that "nine events clearly showed ICME signatures 
in the solar wind observations. However, we were not able to find any 
conventional frontside halo CME candidates in the plausible search window, 
i.e., we fail to identify any eruptive feature on the solar surface 
(e.g., filament eruption, dimming, loop arcade, or long-duration flare), 
in spite of the availability of disk observations from EIT, SXT, or SXI. 
Similar ''problem events'' have been reported earlier 
\citep{Webbetal1998,Zhangetal2003}". 
We think that similar selection methods are used in papers by 
\citet{Gopalswamy2007}
and 
\citet{Zhangetal2007}, because several co-authors took part in both papers.

Thus, we can conclude: 
\begin{enumerate}
\item	Selection method used by 
\citet{Gopalswamy2007} 
is incorrect because it identifies part of frontside halo CMEs (fronside 
halo CMEs without disk activity) as backside halo CMEs;

\item  List of frontside halo CMEs used by 
\citet{Gopalswamy2007} 
is incorrect because it does not include all frontside halo CMEs during 
indicated period; 

\item  Estimation of geoeffectiveness of frontside halo CMEs made by 
\citet{Gopalswamy2007} 
is incorrect because they found a geoeffectiveness only of frontside halo 
CMEs with disk activity.
\end{enumerate}

\begin{acknowledgements} 
       Work was in part supported by RFBR, grant 07-02-00042.  
\end{acknowledgements}

\end{article}


\begin{thebibliography}{}

  \bibitem[{\it Gopalswamy et al.}(2007)]{Gopalswamy2007}
\reference 
Gopalswamy, N., S. Yashiro, and S. Akiyama (2007), Geoeffectiveness of 
halo coronal mass ejections, J. Geophys. Res., 112, A06112, 
doi:10.1029/2006JA012149

\bibitem[{\it Kim et al.}(2005)]{Kimetal2005}
\reference 
Kim, R.-S., K.-S. Cho, Y.-J. Moon, Y.-H. Kim, Y. Yi, M. Dryer, S.-C. 
Bong, and Y.-D. Park (2005), Forecast evaluation of the coronal mass 
ejection (CME) geoeffectiveness using halo CMEs from 1997 to 2003, J. 
Geophys. Res., 110, A11104, doi:10.1029/2005JA011218.

\bibitem[{\it Schwenn et al.}(2005)]{Schwennetal2005}
\reference 
Schwenn, R., A. Dal Lago, E. Huttunen, and W. D. Gonzalez (2005), 
The association of coronal mass ejections with their effects near the 
Earth, Ann. Geophys., 23, 1033-1059. 

\bibitem[{\it Webb et al.}(1998)]{Webbetal1998}
\reference 
Webb, D. F., E.W. Cliver, N. Gopalswamy, H. S. Hudson, and O. C. St. Cyr 
(1998), The solar origin of the January 1997 coronal mass ejection, 
magnetic cloud and geomagnetic storm, Geophys. Res. Lett., 25, 2469- 2472.

\bibitem[{\it Yermolaev et al.}(2005)]{Yermolaevetal2005}
\reference 
Yermolaev Yu. I., M. Yu. Yermolaev, G. N. Zastenker, L.M.Zelenyi, A.A. 
Petrukovich, J.-A. Sauvaud (2005), Statistical studies of geomagnetic storm 
dependencies on solar and interplanetary events: a review, Planetary 
and Space Science, 53/1-3 pp. 189-196.  

\bibitem[{\it Yermolaev and Yermolaev}(2006)]{Yermolaev2006}
\reference 
Yermolaev Yu.I. and M.Yu. Yermolaev (2006), Statistic study on the geomagnetic 
storm effectiveness of solar and interplanetary events, Adv.Space Res., 
37, Issue 6, p. 1175-1181 

\bibitem[{\it Zhang et al.}(2003)]{Zhangetal2003}
\reference 
Zhang, J., K. P. Dere, R. A. Howard, and V. Bothmer (2003), Identification 
of solar sources of major geomagnetic storms between 1996 and 2000, 
Astrophys. J., 582, 520- 533.

\bibitem[{\it Zhang et al.}(2007)]{Zhangetal2007}
\reference 
Zhang, J., et al. (2007), Solar and interplanetary sources of major 
geomagnetic storms (Dst$<$-100 nT) during 1996-2005, J. Geophys. Res., 
112, A10102, doi:10.1029/2007JA012321.

\end{thebibliography}
\end{document}